# THE PASSIVE AND ACTIVE PERIODS FOR THE INTERMITTENT USE OF AN ACTIVE SENSOR TO DETECT AN EVASIVE TARGET

by
Niels Bache
Dec. 2013

Abstract

Your task is to detect a submarine with your active sonar. The submarine can hear your active sonar before you can detect him. If the submarine is fast enough he can evade you before you can detect him. How do you then detect him? If you are using your active sonar continuously you will not detect him. Likewise, if you are not using your sonar at all. In between those two extremes there is an optimum. We will find that optimum. Or said more precisely and general: In the same two dimensional region two platforms are present. One platform, the searcher, equipped with one active sensor (sonar, radar, lidar etc.), is trying to detect the other platform, the target, by means of its active sensor. The target tries to avoid detection using only a passive sensor to detect the searcher. The target can detect the active sensor before the searcher can detect the target (forestalling). The active sensor is therefore used intermittently to surprise the target. The aim of this study is to quantify the passive period of the active sensor by minimizing missed detection opportunities. The active period is subsequently found by maximizing the average detection width of the searcher sensor over time.

# 1. INTRODUCTION

In a two dimensional region one platform, the searcher, equipped with an active sensor, is trying to detect an ideal evasive target present in the same region. The target, in order not to reveal its presence, uses only a passive sensor to detect the searcher. When the searcher is using its active sensor, the target can detect the searcher before the searcher can detect the target (forestalling). The active sensor is therefore used intermittently to surprise the target. The aim of this study is to quantify the passive and active periods of the active sensor, using two measures of effectiveness (MOE) : The passive period is determined by minimizing missed detection opportunities (MOE1). The active period is subsequently found by maximizing the average detection width of the searcher sensor over time (MOE2).

For simplicity we assume straight line courses for searcher and target during the detection process. The duration of the detection process is found in section 2. However, when the target realizes the presence of the searcher, he momentary change his straight line course to maximize the distance to the searcher. If the target speed is greater than the speed of the searcher, it will choose a radial course away from the searcher. If the speed is less than the speed of the searcher, it will choose a course which maximize the lateral range to the searcher (limiting lines of escape).

The active sensor can be any equipment emitting energy and capable of receiving and processing eventual reflected echoes. It can be a sonar, radar, lidar, searchlight or even a torch. The platform can be a war ship or a maritime patrol aircraft (MPA) in AntiSubmarine Warfare (ASW), more seldom a submarine searching for another submarine or surface ship, a fishing boat looking for fish shoals (ref.1), the cost guard searching for smugglers by means of ship or aircraft, war ships looking for pirates or even a truck, equipped with a searchlight, or just a person with a torch, looking for a fleeing person or animal at night in an open field with no hiding places.

In the open literature, many related papers were found, but only a few papers determining the length of the passive and active periods: Ref. 1 treats the fishing boat trying to surprise fish shoals by using the echo sounder intermittently. Mostly about limiting lines of escape. In ref. 2 one frigate in an area is searching for a submarine. The active period is chosen constant and the passive period is a variable. The search route of the frigate is given as is the sensor detection curve versus distance frigate - submarine. The optimum passive period is chosen which maximize the average detection probability over the search route, so the problem is similar but the treatment and assumptions are different. In ref. 3 more than one search unit is possible against a moving target. Near optimum search routes are found by means of a heuristic based on a probability field of position of the target in the area. This field changes over time. At each time step a most promising searcher course is found based on cells in the area with high values of the field. In ref. 4 the analysis is based on data generated by a computer simulation of the relative motion of the searcher and target. The target motion is modeled by a diffusion process. The searcher uses his sensor randomly taken from a distribution. The alerted target can locate the searcher by passive detection of the searcher transmissions. When the target becomes alerted it sprints away from the searcher radially with speed greater than that of the searcher. In ref. 5 the subject is treated as a two person zero-sum dynamic game of the pursuit-evasion type by means of dynamic programming which yields optimal strategies for the searcher and target in a finite state space and in discrete time. When to switch on the searcher sensor is considered a stochastic event determined by the dynamic programming equations. The target can detect the active sensor from everywhere.

# 2. SCENARIO AND DEFINITIONS

We have already outlined the general scenario, but in order to better describe the interaction between searcher and the target, we define

For the **searcher** : S, The range where the searcher can detect the target by means of its active sensor.
V, The speed of the searcher.

For the **target** : R, The range where the target can detect the searcher, when the searcher has his sensor on. S < R.
r, The range where the target can detect the searcher, when the searcher has his sensor off. r < R.
U, The speed of the target.

These five variables are the **input** parameters to the model. We call these variables **the basic parameters**.

We assume an omnidirectional cookie cutter detection model for S, R and r. S can be extended to a two parameter omnidirectional cookie cutter model, see section 9. The model presented here is thus simple but even so, it turns out to be fairly complicated with many cases, see section 10. The model could be considered as a framework.

If S < R is not fulfilled, the target will never be able to detect the searcher before the searcher detects the target. We are then in the continuous use of the active sensor.
If r < R is not fulfilled, there is no point in having a passive period, since the target will detect the searcher even further away than under the active period.

The duration of the detection process: When the active sensor is on, the interaction between searcher and target start, when the distance is R between them and decreasing. When the active sensor is off, the interaction between searcher and target start, when the distance is r between them and decreasing. The interaction stop, when the distance is S and increasing.

Depending of the relative values of r/S and V/U , we divide into various parts, see the summary, section 10, due to different geometries involved.

We call the passive and active period Tp and Ta respectively and the total period T = Tp + Ta .

There are three main **output** : The passive period Tp ,the active period Ta and the maximum average detection width of the searcher over one (or more) periods. The advantage of focusing on the detection width instead of detection probability is that we do not have to consider the size of the region, the searcher route and the search time. It turns out, that without these parameters one can still determine Tp and Ta, according to some chosen criteria (MOE1 and MOE2). Further, the detection width has an operational significance in itself.

We show a numerical example as we go along. The values for this example are
**Searcher :**
$S := 4$
$V := 20$
**Target :**
$R := 8$
$r := 4.5$
$U := 9$
The units for (range, time, speed) could be (nautical mile, hour, knot), (kilometer, hour, km/h), (meters, seconds, m/s) etc. (The sign := is an assignment symbol)

We will look at the target relative to the searcher, i.e. we regard the searcher as **stationary**

and the unalerted target approaching the searcher at a straight line course $\gamma$ relative to searcher heading (the vector **V**) and at a speed $W(\alpha)$ or $W(\gamma)$ ,but we prefer $\gamma$ as the argument , see the figure 1 below.

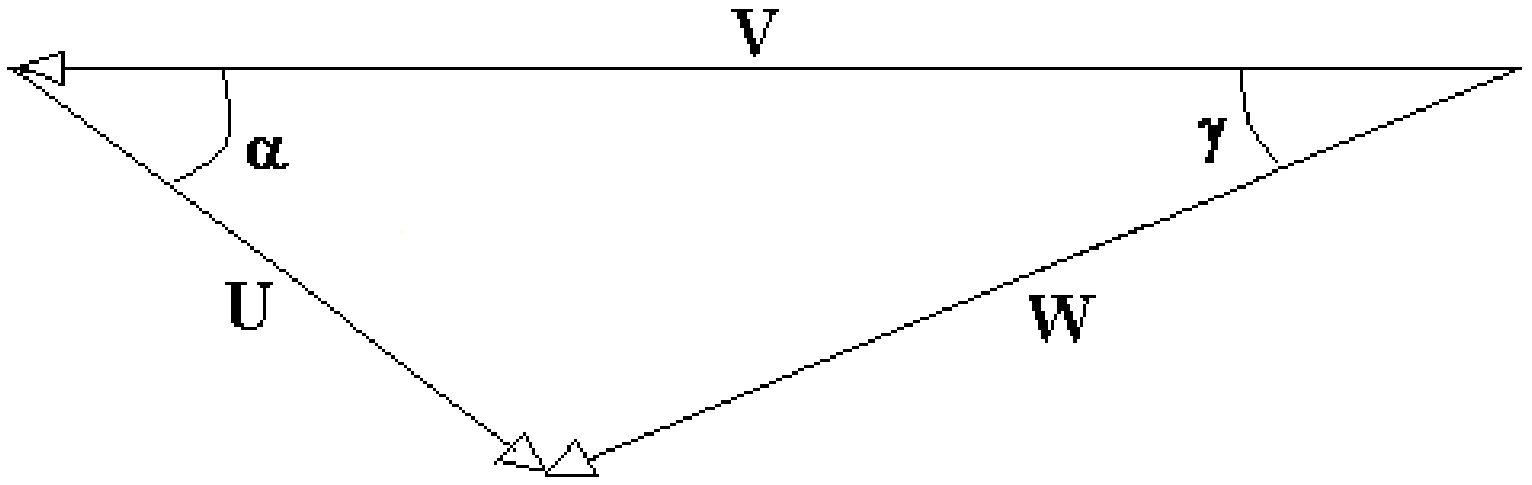

**Figure 1. Target and searcher velocities.**

$$W(\alpha) := \sqrt{U^2 + V^2 - 2 \cdot U \cdot V \cdot \cos(\alpha)}$$

We assume symmetry with respect to the vector **V**. Before the target is being alerted to the presence of the searcher, we assume $\alpha$ to be a random variable, $0 \le \alpha \le \pi$ , uniformly distributed in lack of any knowledge of $\alpha$ . The minimum and maximum values for $W(\alpha)$ are V - U and V + U respectively and the optimum evasive speed is $\sqrt{V^2 - U^2}$ ,for $U \le V$ and U - V for $V \le U$ (away from the searcher).

The mean value is

$$\text{meanW} := \frac{1}{\pi} \cdot \int_0^{\pi} W(\alpha)\, d\alpha$$

The example : meanW = 21.0

# 3. INSTANTANEOUS DETECTION WIDTH AND CONDITION FOR DETECTION

In this section and until further notice we assume

$$U < V \text{ and } S < r$$

If V < U and still S < r there will be no possibility for detection as we have assumed, that the target will flee radially away from the searcher before the searcher can detect the target (S < r). The only hope for detection when V < U is, when r < S, see section 7.

**Limiting lines of escape for evasive targets:**

Limiting lines of escape are two tangents to the circle S, see Fig. 2 below, within which the submarine can not escape detection due to lack of speed. We need the two tangents to determine the detection width D1 when the submarine become alerted of the presence of the searcher.

We assume, that the target, when alerted, adopt the best evasive absolute course $\alpha_e$ to avoid the searcher (worst case from the searcher point of view), see figure 2a below, corresponding to

$$W(\alpha_e) = \sqrt{V^2 - U^2}$$

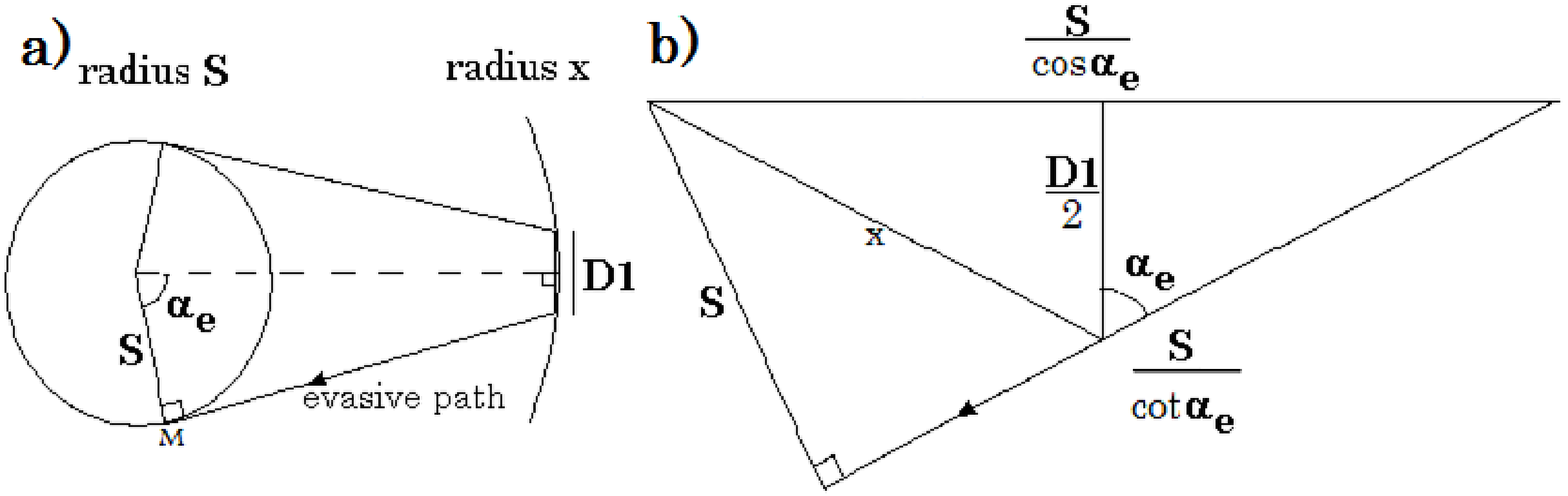

**Figure 2. Limiting lines of escape.**

We have from figure 2b, x can be either R or r :

$$\cos\alpha_e = \frac{U}{V}$$

and

$$\frac{\frac{D1}{2}}{S} = \frac{\frac{S}{\cot\alpha_e} - \sqrt{x^2 - S^2}}{\frac{S}{\cos\alpha_e}}$$

, giving

$$\alpha_e := \operatorname{acos}\left(\frac{U}{V}\right)$$

$$\alpha_e = 63 \cdot \text{deg}$$

and the chord D1, interpreted as **the instantaneous detection width** (in contrast to the later maximum average detection width).

$$D1(x) := \begin{cases} 2\cdot S\cdot\left[\sqrt{1-\left(\frac{U}{V}\right)^2} - \frac{U}{V}\cdot\sqrt{\left(\frac{x}{S}\right)^2 - 1}\right] & \text{if } x < S\cdot\frac{V}{U} \\ 0 & \text{otherwise} \end{cases}$$

(1)

D1(x) is the instantaneous detection width, when the target becomes alerted at a distance x from the target. For the example we have

$D1(r) = 5.29$ $\quad$ $D1(R) = 0.91$

$S\cdot\frac{V}{U} = 8.89$

D1(R) is the detection width in the continuous case.

$$S\cdot\frac{V}{U}$$

is the maximum distance from the sensor, where there still is a positive detection width.

If the approaching unalerted target touches D1, it will be detected whatever course it chooses, provided $U < V$ and provided the sensor is switched on before the target reaches the point M or leaves the circle S, figure 2a, otherwise it will be a missed opportunity.

The condition in Eq.(1) give $r < S\cdot V/U$ or $R < S\cdot V/U$ . If the latter is valid, the former is also valid ($r < R$), but no vice versa, i.e. either we have a positive detection width on both circles r and R or just on the circle r. If $D1(r) = 0$, there will be no detection at all, not even with the intermittent use of the sensor. This Leads to the **condition for detection**

$$\frac{U}{V} < \frac{S}{r}$$

(2)

# 6. Visualization of the intermittent case

Our scenario is basically a 2-dimensional problem (searcher and targets in the X,Y plane). As the searcher switch its sensor on and off, time must play a crucial role. In order to visualize time, we choose time as the third ax upwards (Z direction) in 3 dimensions. The on- and off periods (Ta and Tp) can then be pictured as vertical cylinders with radii R and r in the figures below. We start at time $t = 0$, see figure 3b, at the beginning of the passive period and at a range R from the searcher, where a target, if present, is approaching the searcher. The target approach the searcher at a positive slope relative to the X,Y plane with angle acot(W(α)). The aim is to visualize Tp ,Ta ,R ,r and D1(x) versus time.

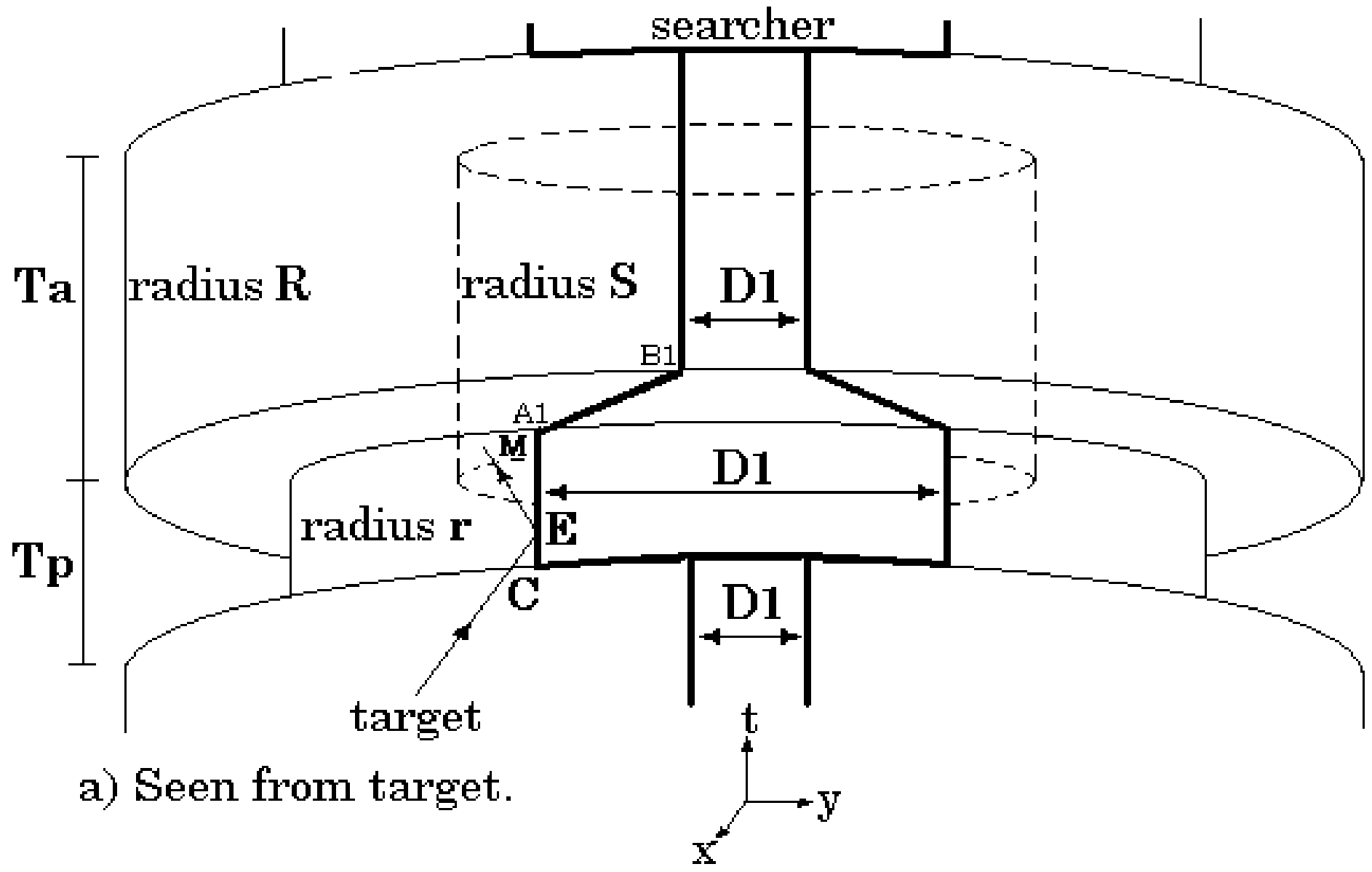

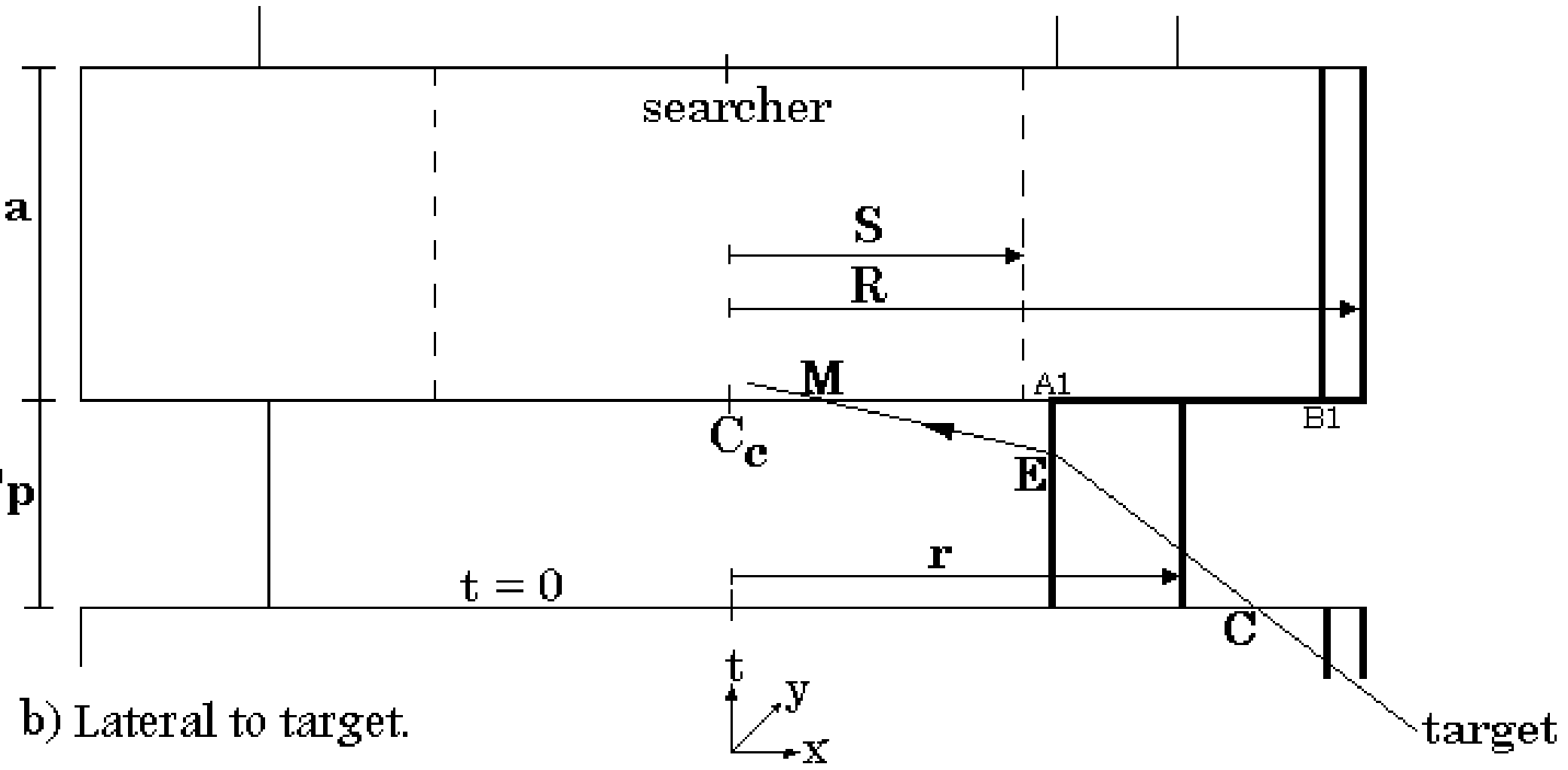

**Figure 3. Visualization of variables.**

# 5. THE PASSIVE PERIOD

We are free to choose the passive period Tp. With a short Tp we approach the continuous case. With a long Tp we will rarely make a detection. Tp should be sufficient large to accumulate opportunities, if they exist, see Ineq.(2). On the other hand, if the main mission - or one of the important missions - of the platform is to detect targets, these opportunities should not become missed opportunities. We adopt therefore the following criteria, measure of effectiveness ( MOE), for determining Tp :

MOE1: Opportunities created during the passive period Tp must not become missed opportunities due to too long a Tp.

Tp should then be equal to the minimum time it takes for a target to present an opportunity and then leave the detection circle S, i.e. become a missed opportunity. This gives a value of Tp .

We distinguish between the target being alerted or unalerted. If the target is unalerted (outside r), the path is CE for a given , see figure 1. If the target is alerted (being inside the circle r, see figure 4 below), the quickest path is EM, the limiting line of escape, which we assume the target will adopt.

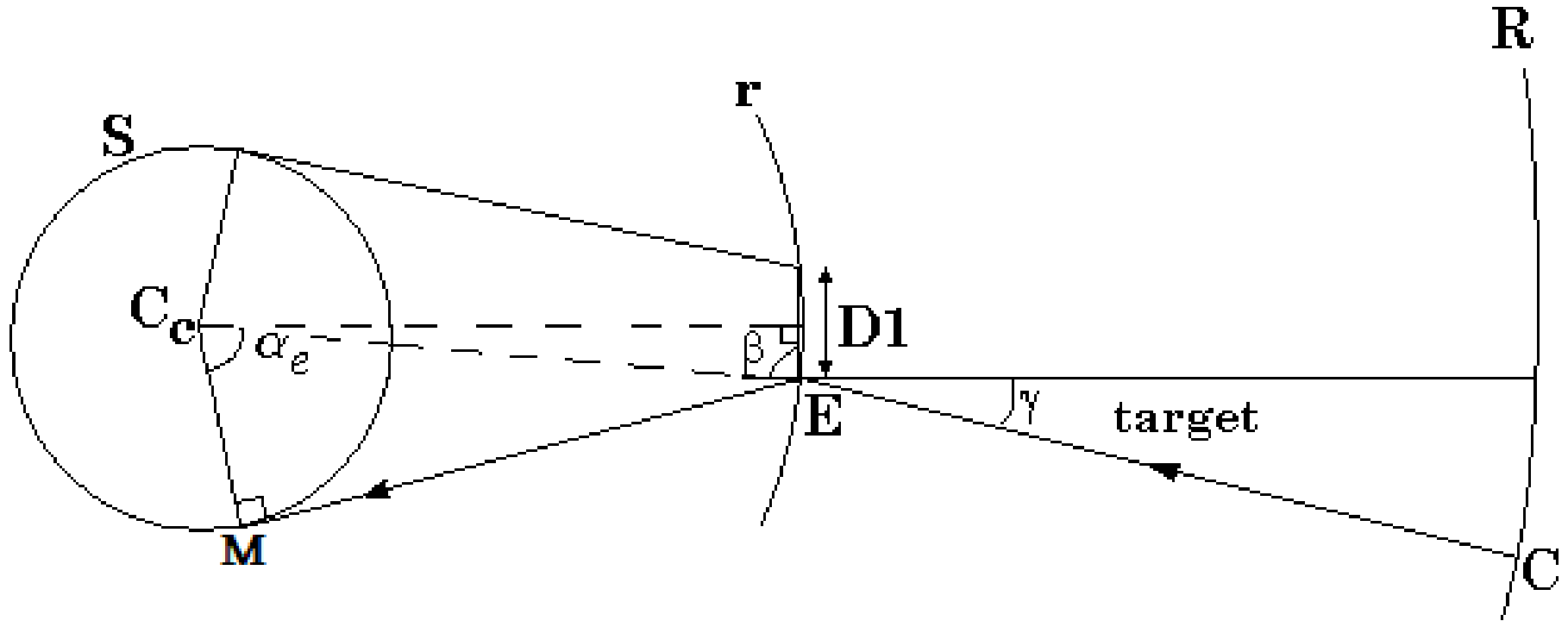

**Figure 4. Quickest target path to become a missed detection opportunity.**

We have from figure 1 and 4

$$\beta(r) := \operatorname{acos}\left(\frac{D1(r)}{2\cdot r}\right)$$

and

$$\gamma(\alpha) := \operatorname{asin}\left(\frac{U}{W(\alpha)}\cdot\sin(\alpha)\right)$$

$$R^2 = r^2 + CE^2 - 2\cdot r\cdot CE\cdot\cos\left(\frac{3\cdot\pi}{2} - \beta - \gamma\right)$$

(figured 4, triangle $CEC_c$)

$$CE = r\cdot\cos\left(\frac{3\cdot\pi}{2} - \beta(r) - \gamma\right) + \sqrt{\left(r\cdot\cos\left(\frac{3\cdot\pi}{2} - \beta(r) - \gamma\right)\right)^2 + R^2 - r^2}$$

or

$$CE(\alpha) := \begin{cases} \sqrt{R^2 - r^2\cdot\cos(\beta(r) + \gamma(\alpha))^2} - r\cdot\sin(\beta(r) + \gamma(\alpha)) & \text{if } \frac{U}{V} < \frac{S}{r} \\ 0 & \text{otherwise} \end{cases}$$

(3)

The time for the target to travel the distance EM (figure 4) is

$$\frac{\sqrt{r^2 - S^2}}{\sqrt{V^2 - U^2}}$$ ,if Ineq.(2) holds.

**The passive period Tp** is then the time it takes for the target to steam the distance CEM :

$$Tp(\alpha) := \frac{CE(\alpha)}{W(\alpha)} + \left(\frac{U}{V} < \frac{S}{r}\right)\cdot\frac{\sqrt{r^2 - S^2}}{\sqrt{V^2 - U^2}}$$

(4a)

With this value of Tp and a given target course a, we have now ensured MOE1, provided the target change to the evasive course when alerted. Sometime one has a clue, more or less accurate, of the unalerted target course a , therefore we keep as a parameter.

The example :

$\alpha := 0$

$Tp(\alpha) = 0.47$

$\alpha := \pi$

$Tp(\alpha) = 0.25$

The mean value of Tp :

$$meanTp := \frac{1}{\pi}\cdot\int_0^{\pi} Tp(\alpha)\, d\alpha$$

$meanTp = 0.31$

, assuming uniformly distributed , i.e. no knowledge of $\alpha$ .

If meanTp is used instead of Tp *,* we can no longer ensure MOE1 due to the assumed randomness. If we insist on MOE1, we should use the minimum value of Tp, if we have no knowledge of a :

$\alpha min := Minimize(Tp, \alpha)$

$\alpha min = 2.74$

radians

$Tp(\alpha min) = 0.25$

In the following, both the mean value and the minimum value of Tp is given .

# 6. THE ACTIVE PERIOD. DETECTION WIDTH VERSUS TIME

When introducing a passive period - the blind period - we will of cause loose eventual opportunities from the immediate past. On the other hand, by means of the passive period, we will allow the target to eventually come closer before being alerted, increasing the detection probability, i.e. increasing the instantaneous detection width D(t). The question is, whether the loss of detection width (D = 0) during Tp will be outweighed by the gain in detection width later. More precisely, we will draw D(t) versus time t and if the *average* D is greater than D1(R) it is advantageous to use the intermittent case, otherwise we should stick to the traditional continuous case. If D1(R) = 0 however, as often is the case, we have to use the intermittent case to get detection's at all . We can draw a generic picture of the instantaneous detection width D versus time, figure 5 below.

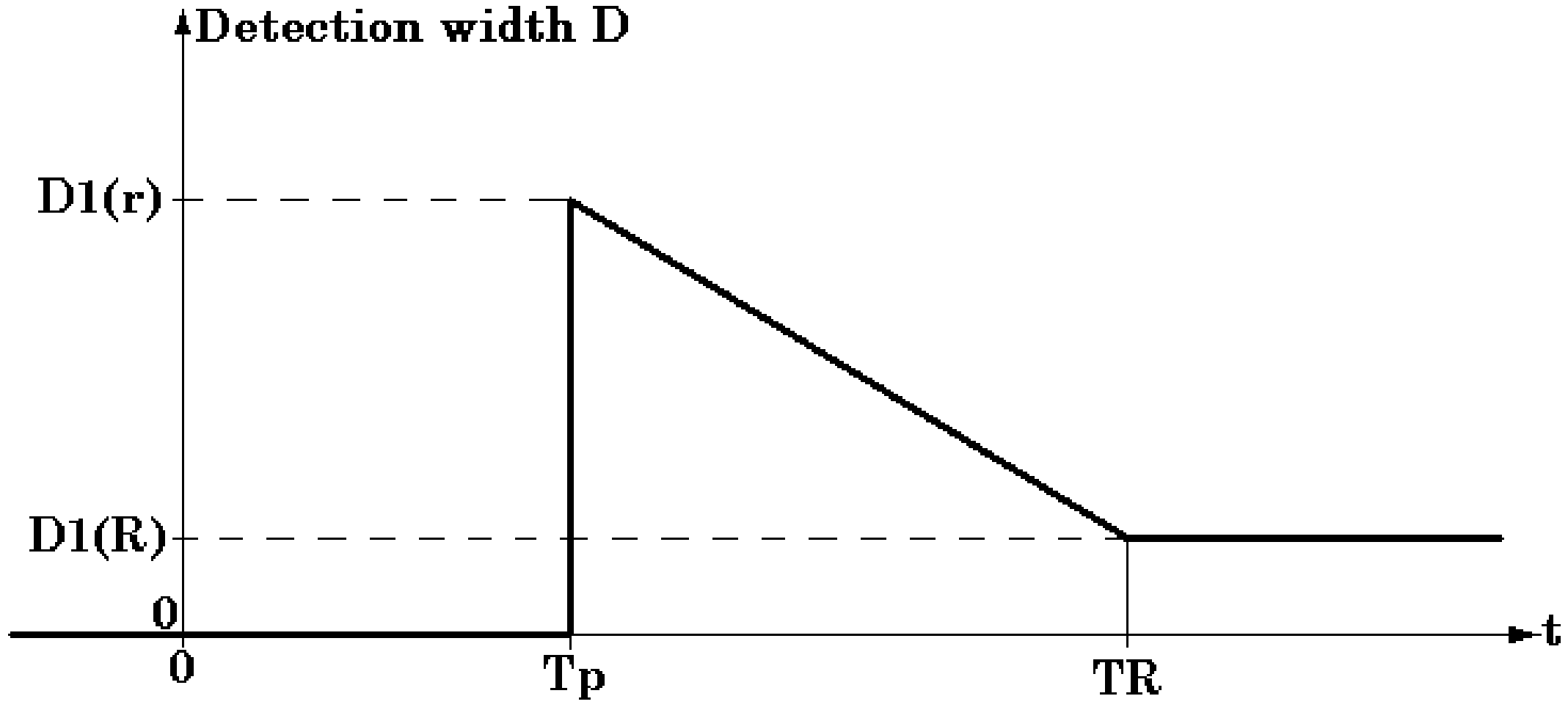

Figure 5. Detection width versus time. Generic.

During the passive period detection opportunities have accumulated between the circle R and circle r. At the beginning of the active period all these targets - now alerted - takes the evasive course $\alpha_e$ and they are faced with the detection width D1(r) . After the time $CE(\alpha_e)/W(\alpha_e)$ all these targets have either been inside the limiting lines of escape, with the possibility of detection, or outside and we are back to the detection width D1(R) , therefore

$$TR(\alpha) := Tp(\alpha) + \frac{CE(\alpha_e)}{W(\alpha_e)}$$

(5a)

Example :

$\alpha := 0$

$TR(\alpha) = 0.67$

$\alpha := \pi$

$TR(\alpha) = 0.45$

Between time Tp and TR the range argument x in Eq.(1) increases linearly with time from r to R (figure 5) as

$$x(\alpha,t) := \begin{cases} \frac{TR(\alpha) - t}{TR(\alpha) - Tp(\alpha)} \cdot r + \frac{t - Tp(\alpha)}{TR(\alpha) - Tp(\alpha)} \cdot R & \text{if } R < S \cdot \frac{V}{U} \\ \frac{TR(\alpha) - t}{TR(\alpha) - Tp(\alpha)} \cdot r + \frac{t - Tp(\alpha)}{TR(\alpha) - Tp(\alpha)} \cdot S \cdot \frac{V}{U} & \text{otherwise} \end{cases}$$

(6a)

From figure 5, we then have the instantaneous detection width D(α,t)

$$D(\alpha,t) := \begin{cases} 0 & \text{if } 0 \le t \wedge t < Tp(\alpha) \\ D1(x(\alpha,t)) & \text{if } Tp(\alpha) \le t \wedge t < TR(\alpha) \\ D1(R) & \text{if } TR(\alpha) \le t \end{cases}$$

(7a)

Example :

$\alpha := \frac{\pi}{2}$

and variation of t as

$t := 0, 0.0004 .. TR(\alpha) + 0.1$

gives the graph :

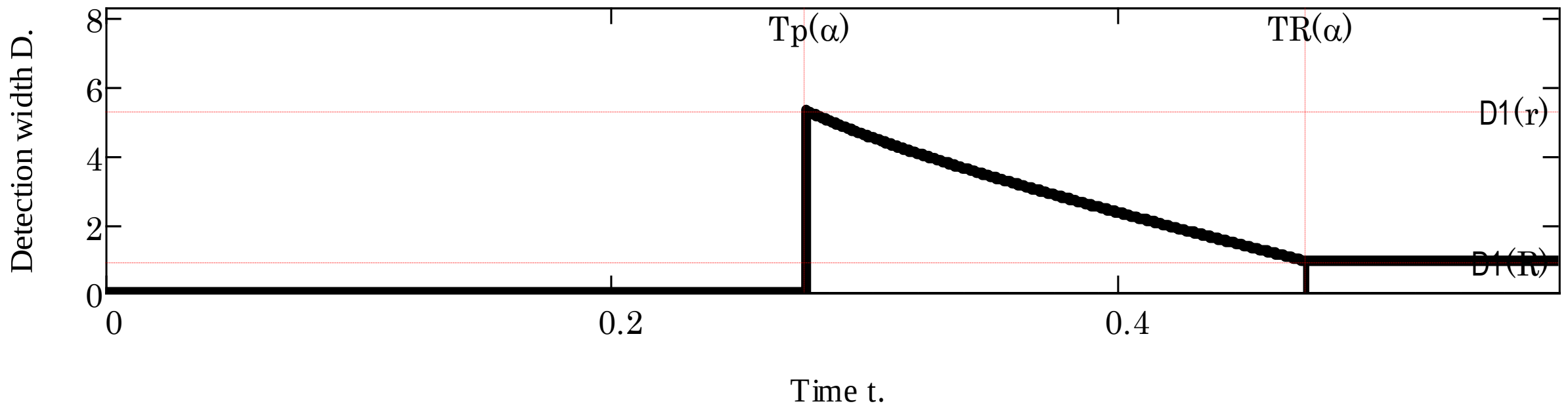

**Figure 6a. Detection width versus time t. Actual.**

The area under the detection width is

$$A(\alpha,t) = \int_0^t D(\alpha,\tau)\,d\tau$$

By means of **Appendix A ,** A(α,t) can be found without the integration sign. The average detection width A(α,t)/t can be displayed :

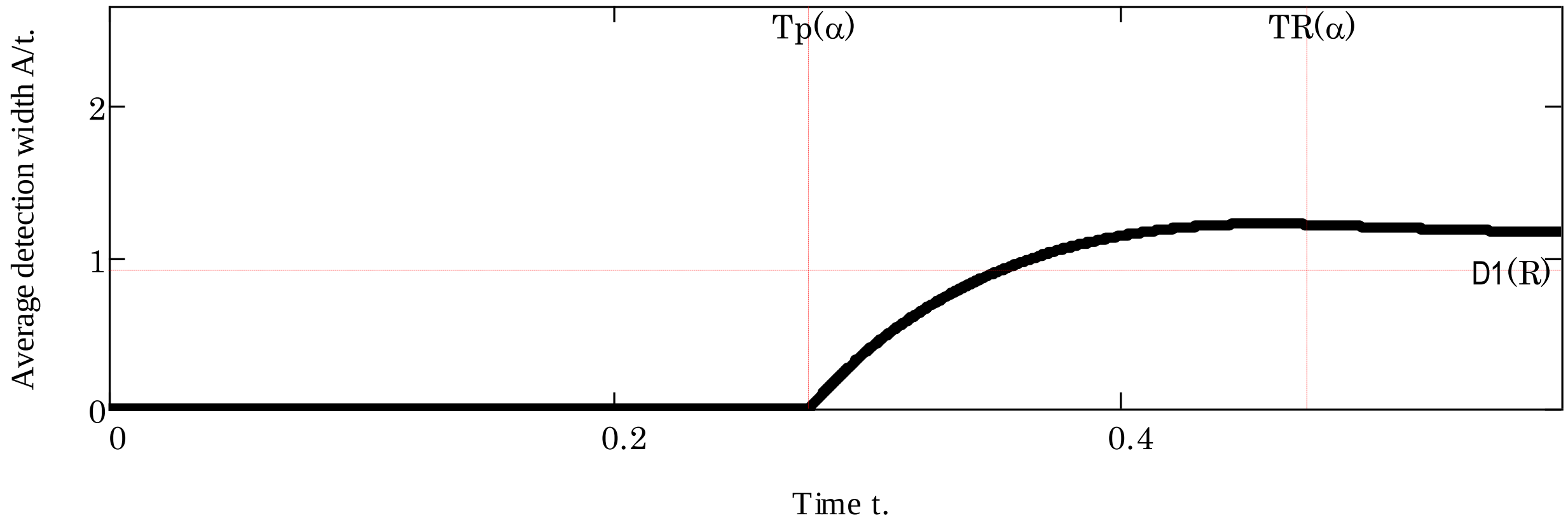

**Figure 7a. Average detection width A/t versus time t. Actual.**

We choose the total period T =Tp + Ta according to the following measure of effectiveness :

MOE2 : Choose the total period T as the time, where the average detection width reach its maximum. If a maximum does not exist, use the continuos case.

For this example we have a maximum. It a maximum does not exist, A(α,t)/t will lie below D1(R) as A(,t)/t approach D1(R) when t becomes large (the continuous case). If a maximum exist, it will lie between Tp(α) and TR(α) due to the shape of D(α,t) , figure 6a.

To find the maximum of the curves in figure 7a, we differentiate A(α,t)/t with respect to the time interval Tp to TR and set it equal to zero, giving

$$t \cdot D(\alpha,t) - A(\alpha,t) = 0$$

(8)

, provided D is a continuous function of t and A is a differential function of t between Tp < t < TR , which the are.

We define

$y(\alpha,t) := t \cdot D(\alpha,t) - A(\alpha,t)$
(8a)

An explicit solution for t in Eq. (8) does not exist, since I(x) , which is part of A(α,t) see Appendix A at the end, contains the ln function, i.e. Eq. (8) is a transcendental equation. However, y(α,t) is almost a linear function in t as shown in the figure 8a below. An approximation to y(α,t) would then be the function z(α,t) :

$$z(\alpha,t) := y(\alpha,Tp(\alpha)) + \frac{y(\alpha,TR(\alpha)) - y(\alpha,Tp(\alpha))}{TR(\alpha) - Tp(\alpha)} \cdot (t - Tp(\alpha))$$

(9a)

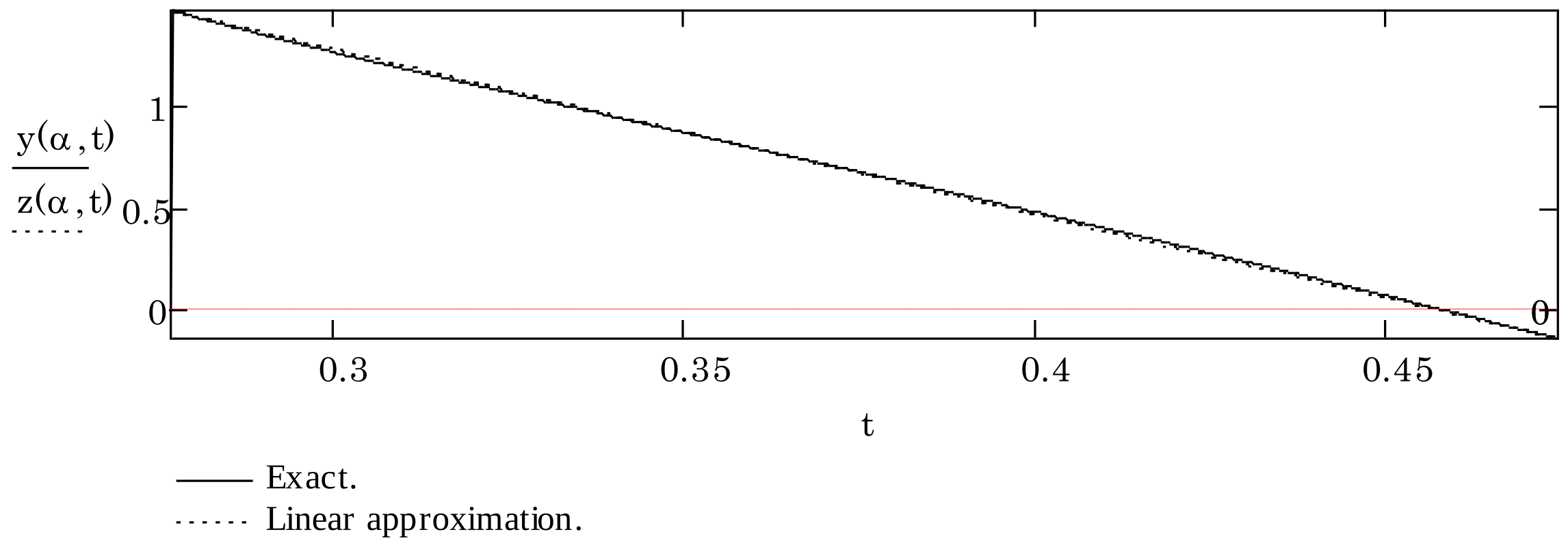

**Figure 8a. Solution for *T*.**

It is now easy to find a good approximation to T, when setting *z*(α,t) = 0 in Eq. (9a) :

$$T(\alpha) := Tp(\alpha) - \frac{TR(\alpha) - Tp(\alpha)}{y(\alpha,TR(\alpha)) - y(\alpha,Tp(\alpha))} \cdot y(\alpha,Tp(\alpha))$$

and therefore

$Ta(\alpha) := T(\alpha) - Tp(\alpha)$
(10a)

$\alpha := 0$

$Ta(\alpha) = 0.2$

$T(\alpha) = 0.67$

$\alpha := \pi$

$Ta(\alpha) = 0.18$

$T(\alpha) = 0.43$

with the mean values

$$\frac{1}{\pi} \cdot \int_0^{\pi} Ta(\alpha)\, d\alpha = 0.18$$

$$meanT := \frac{1}{\pi} \cdot \int_0^{\pi} T(\alpha)\, d\alpha$$

$meanT = 0.49$

, assuming a uniformly distributed .

If we define the dimensionless quantity

$$\rho(\alpha) := \frac{-y(\alpha,Tp(\alpha))}{y(\alpha,TR(\alpha)) - y(\alpha,Tp(\alpha))}$$

,we have another expression for the optimum **total period *T***

$T(\alpha) := Tp(\alpha) + \rho(\alpha)\cdot(TR(\alpha) - Tp(\alpha))$

Another criteria for adopting an intermittent sensor policy is then
$0 < \rho(\alpha) < 1$

The example:
$\alpha := 0$
$\rho(\alpha) = 1.01$
, corresponding to having no maximum in figure 7a
$\alpha := \pi$
$\rho(\alpha) = 0.89$
$\alpha := 0, 0.01 .. \pi$
gives the graph :

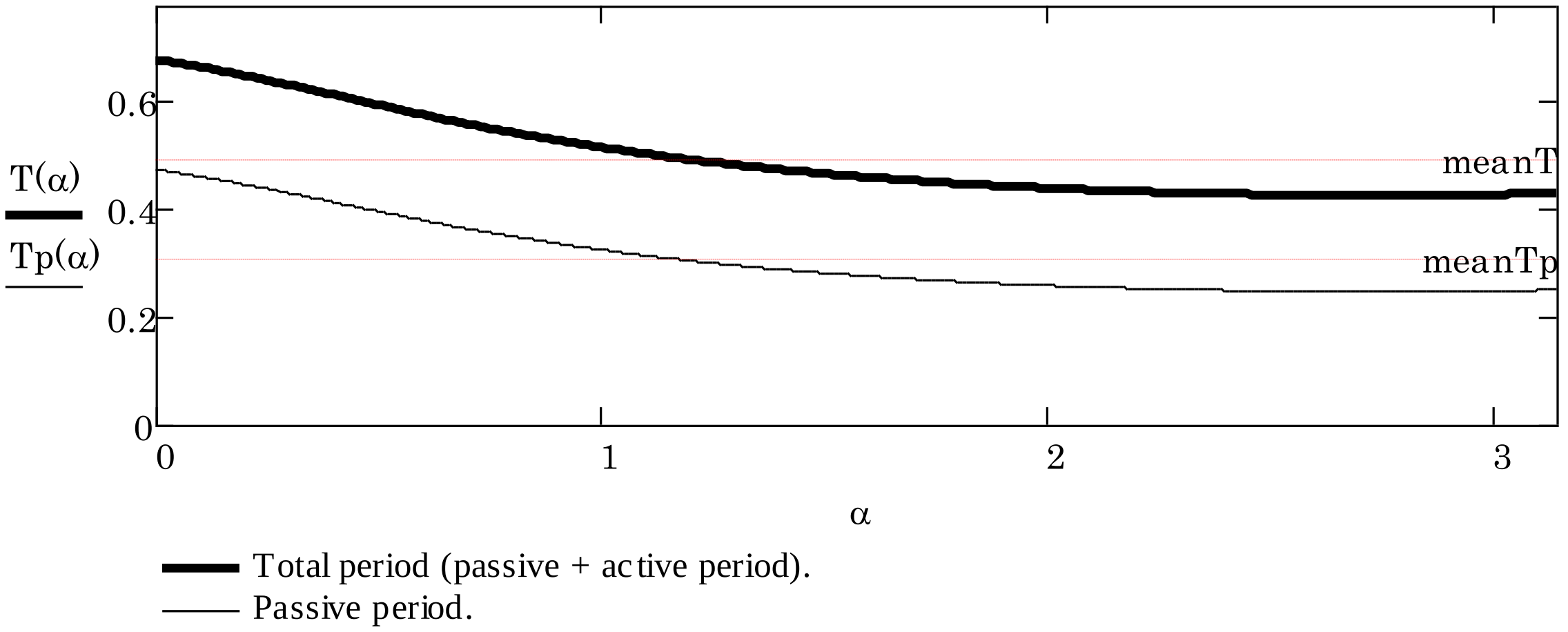

**Figure 9. Passive and total period.**

$$meanT := \frac{1}{\pi}\cdot\int_0^{\pi} T(\alpha)\, d\alpha$$

meanT = 0.49

**The maximum average detection width** (the third **output** parameter) for the example is

$\alpha := 0$

$$\frac{A(\alpha, T(\alpha))}{T(\alpha)} = 0.86$$

$\alpha := \pi$

$$\frac{A(\alpha, T(\alpha))}{T(\alpha)} = 1.29$$

The mean value with respect to   is

$$meanD := \frac{1}{\pi}\cdot\int_0^{\pi} \frac{A(\alpha, T(\alpha))}{T(\alpha)}\, d\alpha$$

meanD = 1.16

compared to the continuous case :

D1(R) = 0.91

**The Gain** *G* compared to the continuous case is then on the average

$$G := \begin{cases} \frac{meanD}{D1(R)} - 1 & \text{if } 0 < D1(R) \\ 100 & \text{otherwise} \end{cases}$$

$G = 27\cdot\%$

**The maximum average detection area per time unit** is

$$\frac{1}{\pi}\cdot\int_0^{\pi} W(\alpha)\cdot\frac{A(\alpha, T(\alpha))}{T(\alpha)}\, d\alpha = 25$$

compared to the approximation

meanW·meanD = 24

## 6.1 Limited time for new opportunities

Until now we have assumed that

$$R < S \cdot \frac{V}{U}$$

, see Eq.(1) and the example.

This means that during the whole active period opportunities might continuously emerge as $0 < D1(R)$ :
$D1(R) = 0.91$
If

$$S \cdot \frac{V}{U} \le R$$

then new opportunities will stop emerging after the time

$$\frac{\sqrt{\left(S \cdot \frac{V}{U}\right)^2 - r^2 \cdot \cos(\beta(r) + \gamma(\alpha))^2} - r \cdot \sin(\beta(r) + \gamma(\alpha))}{\sqrt{V^2 - U^2}}$$

from the beginning of the active period, the nominator is $CE(\alpha_e)$ with $R = S \cdot V/U$ .
After this time there is no point in continuing the active period.

# 7. TARGET ALERT RANGE IN THE PASSIVE PERIOD LESS THEN SEARCHER DETECTION RANGE

The new assumption is r < S , as notified in section 3. We assume, as in section 5, that the target when alerted, adopt the best evasive course (ideal target). the treatment in the following sections is similar to the previous sections, therefore we do it short. On the other hand, this section is more complicated than the previous section, because two cases emerge.
We change the example to
$r := 3.5$
, all other values are the same.

## 7.1 The passive period

As in section 5, Tp should be equal to the minimum time it takes for a target to present an opportunity and then leave the detection circle S, i.e. become a missed opportunity (MOE1).

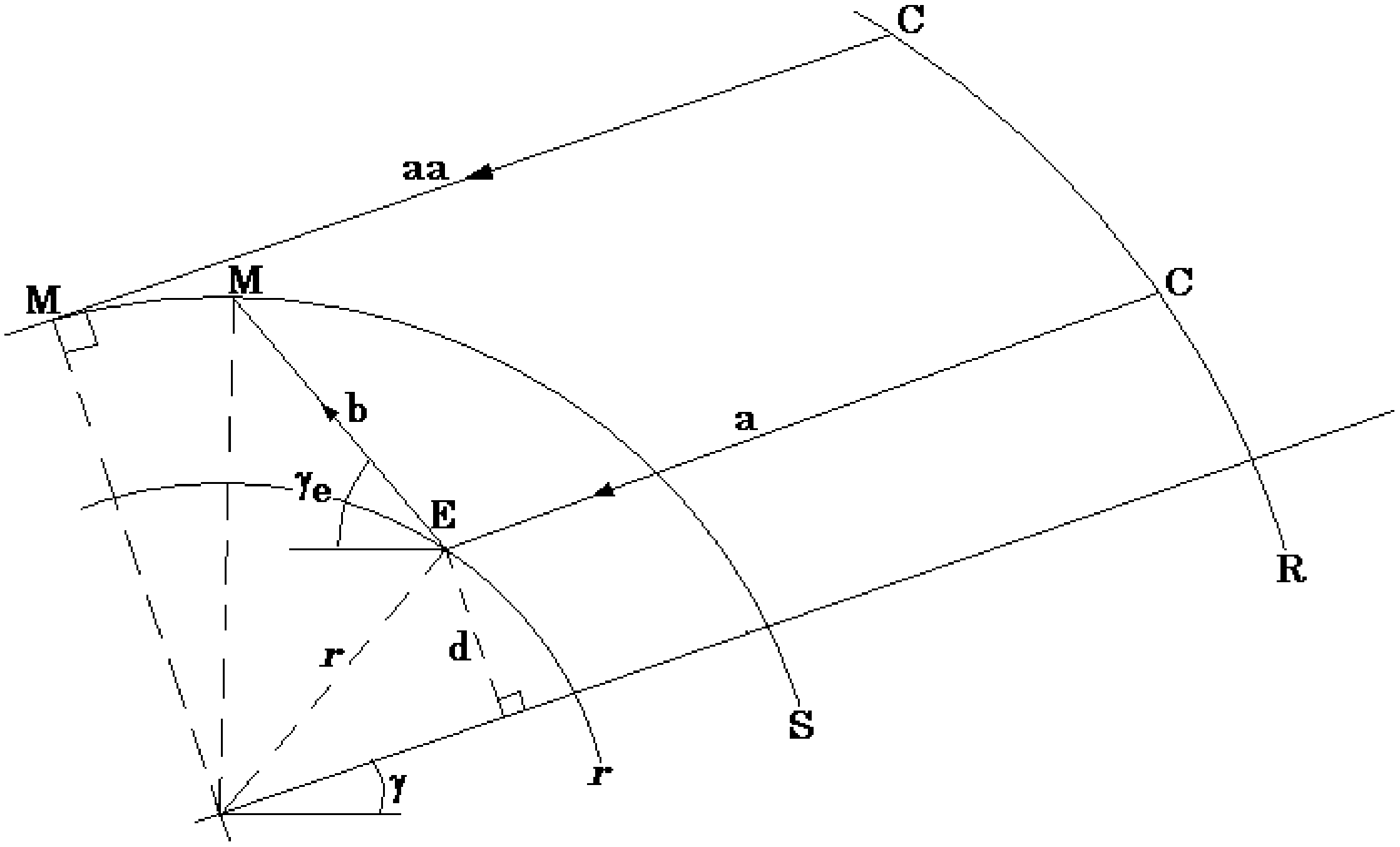

**Figure 10. Target paths for r < S .**

We will find the time T1 and T2 from the target being at the circle R to the time the target ceases as an opportunity, and then choose the minimum of T1 as T2, again in order not to loose opportunities created during Tp, according to MOE1. .

In figure 10 two cases appear :

1) The target course intercepts the circle r at E, i.e. the target becomes alerted, and from thereon the target takes an evasive course $\alpha_e$ , corresponding to $\gamma_e$ (figure 1) or
2) The target goes free of the circle r, i.e. it remains unalerted, but hits the circle S two times.

The lateral range target-searcher d in figure 10 is a parameter, which determines the position C, together with $\alpha$ (or $\gamma$). $0 < d < S$.

Case 1) :

$$T1 = \frac{a}{W(\alpha)} + \frac{b}{W(\alpha_e)}$$

if "a" (the track of the target) hits the circle r (d < r), otherwise

Case 2) :

$$T2 = \frac{aa}{W(\alpha)}$$

where "aa" (the track of the target) is outside the circle r (r < d), but intersects the circle S two times. It is seen that T2 has minimum for a given when aa just touches the circle S. Therefore we substitute T2 with

$$T2(\alpha) := \frac{\sqrt{R^2 - S^2}}{W(\alpha)}$$

If r is close to zero, we have T2 < T1 . If r is close to S, we have T1 < T2 .

$$a(d) := \sqrt{r^2 + R^2 - 2\cdot r\cdot R\cdot \cos\left(\operatorname{asin}\left(\frac{d}{r}\right) - \operatorname{asin}\left(\frac{d}{R}\right)\right)}$$

if $d \le r$

(11)

$$b(\alpha, d) := r\cdot\cos\left(\gamma\left(\alpha_e\right) + \gamma(\alpha) + \operatorname{asin}\left(\frac{d}{r}\right)\right) + \sqrt{S^2 - r^2\cdot\sin\left(\gamma\left(\alpha_e\right) + \gamma(\alpha) + \operatorname{asin}\left(\frac{d}{r}\right)\right)^2} \qquad (12a)$$

$$T1(\alpha, d) := \frac{a(d)}{W(\alpha)} + \frac{b(\alpha, d)}{\sqrt{V^2 - U^2}}$$

Tp with parameters and d is then

$$Tp\alpha d(\alpha, d) := \begin{cases} \dfrac{\sqrt{R^2 - S^2}}{W(\alpha)} & \text{if } \dfrac{\sqrt{R^2 - S^2}}{W(\alpha)} \le \dfrac{a(d)}{W(\alpha)} + \dfrac{b(\alpha, d)}{\sqrt{V^2 - U^2}} \\ \dfrac{a(d)}{W(\alpha)} + \dfrac{b(\alpha, d)}{\sqrt{V^2 - U^2}} & \text{otherwise} \end{cases}$$

, which should be used, if we have some knowledge of d and $\alpha$ .

If we have no information about d :
To fulfill MOE1 we will use the minimum of Tpαd(α,d)d with respect to d and still keep α as a parameter :

$$mind(\alpha) := Minimize(Tp\alpha d, d)$$

giving thus a minimum of Tp with respect to d :

$$Tp(\alpha) := \begin{cases} \dfrac{\sqrt{R^2 - S^2}}{W(\alpha)} & \text{if } \dfrac{\sqrt{R^2 - S^2}}{W(\alpha)} \le \dfrac{a(mind(\alpha))}{W(\alpha)} + \dfrac{b(\alpha, mind(\alpha))}{\sqrt{V^2 - U^2}} \\ \dfrac{a(mind(\alpha))}{W(\alpha)} + \dfrac{b(\alpha, mind(\alpha))}{\sqrt{V^2 - U^2}} & \text{otherwise} \end{cases}$$

(4b)

, which we prefer if we have some knowledge of for example a general target approach direction.

If we have no such knowledge of α and average Tp over α , MOE1 can no longer be ensured due to the assumed randomness :
Example :

$$meanTp := \frac{1}{\pi}\cdot\int_0^{\pi} Tp(\alpha)\, d\alpha$$

$$meanTp = 0.36$$

Instead, we prefer the absolute minimum of Tp :

$$\begin{pmatrix} \alpha min \\ dmin \end{pmatrix} := Minimize(Tp\alpha d, \alpha, d)$$

$$Tp\alpha d(\alpha min, dmin) = 0.24$$

## 7.2 The active period. Detection width versus time

The time from the start of the active period t= Tp(α) until all opportunities - now alerted - created during the passive period have reached the circle S but not yet left the circle, i.e. no missed opportunities (MOE1), is

$$\frac{\sqrt{R^2 - S^2}}{\sqrt{V^2 - U^2}}$$

, see the distance EM in Figure 4 with r replaced with R. After that time the detection width is back to D1(R).

$$TR(\alpha) := Tp(\alpha) + \frac{\sqrt{R^2 - S^2}}{\sqrt{V^2 - U^2}}$$

For Tp < t the range argument x in Eq.(1) increases linearly with time as

$$x(\alpha,t) := \begin{cases} \dfrac{TR(\alpha)-t}{TR(\alpha)-Tp(\alpha)}\cdot S + \dfrac{t-Tp(\alpha)}{TR(\alpha)-Tp(\alpha)}\cdot R & \text{if } R < S\cdot\dfrac{V}{U} \\ \dfrac{TR(\alpha)-t}{TR(\alpha)-Tp(\alpha)}\cdot S + \dfrac{t-Tp(\alpha)}{TR(\alpha)-Tp(\alpha)}\cdot S\cdot\dfrac{V}{U} & \text{otherwise} \end{cases}$$

(6b)

At t = Tp the target is still unalerted. A moment later, Tp < t , the target becomes alerted due to R and the instantaneous detection width start with the value $D1(S) = 2\cdot S\cdot\sqrt{1-\left(\frac{U}{V}\right)^2}$ , which is less than 2S .

$$D(\alpha,t) := \begin{cases} 0 & \text{if } 0 \le t \wedge t < Tp(\alpha) \\ D1(S) & \text{if } t = Tp(\alpha) \\ D1(x(\alpha,t)) & \text{if } Tp(\alpha) < t \wedge t < TR(\alpha) \\ D1(x(\alpha,TR(\alpha))) & \text{if } TR(\alpha) \le t \end{cases}$$

(7b)

Example :

$\alpha := \frac{\pi}{2}$

$t := 0, 0.0008 .. TR(\alpha) + 0.1$

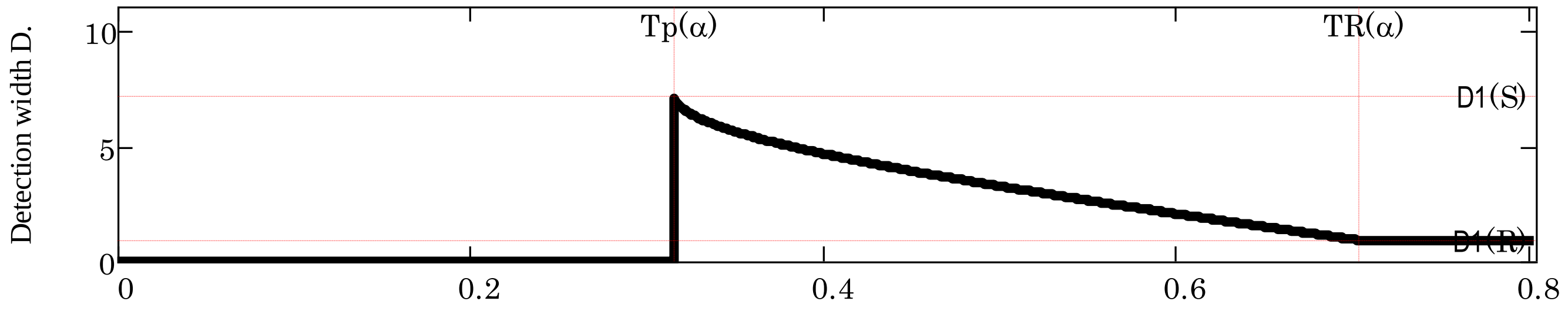

**Figure 6b. Detection width versus time t for r < S . Actual.**

The discontinuity of D at t = Tp has no effect on the area under the curve, figure 6b. Therefore

$$A1(\alpha,t) := \int_{Tp(\alpha)}^{t} D1(x(\alpha,\tau))\,d\tau$$

or (Appendix A)

$$A1(\alpha,t) := \begin{cases} 2\cdot\dfrac{TR(\alpha)-Tp(\alpha)}{R-r}\cdot(I(x(\alpha,t))-I(S)) & \text{if } R < S\cdot\dfrac{V}{U} \\ 2\cdot\dfrac{TR(\alpha)-Tp(\alpha)}{S\cdot\dfrac{V}{U}-r}\cdot(I(x(\alpha,t))-I(S)) & \text{otherwise} \end{cases}$$

$$A(\alpha,t) := \begin{cases} 0 & \text{if } 0 \le t \wedge t < Tp(\alpha) \\ A1(\alpha,t) & \text{if } Tp(\alpha) \le t \wedge t < TR(\alpha) \\ A1(\alpha, TR(\alpha)) + D1(x(\alpha, TR(\alpha)))\cdot(t - TR(\alpha)) & \text{if } TR(\alpha) \le t \end{cases}$$

Example :

$t := 0, 0.005 .. TR(\alpha) + 0.1$

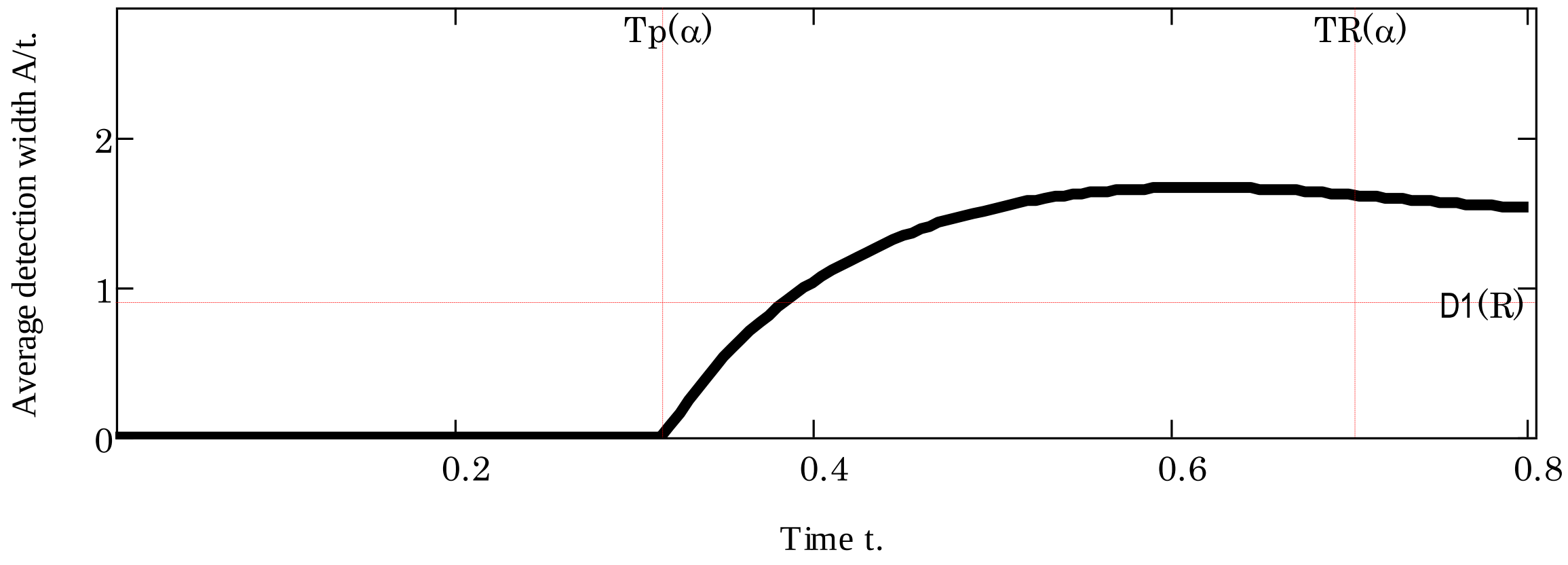

**Figure 7b. Average detection width A/t versus time t for r < S . Actual.**

Since the discontinuity of D at t = Tp has no effect on the area, it has no effect either on finding T, as T is only a function of the area A under the curve D. Therefore we set

$$D(\alpha,t) := \begin{cases} 0 & \text{if } 0 \le t \wedge t < Tp(\alpha) \\ D1(x(\alpha,t)) & \text{if } Tp(\alpha) \le t \wedge t < TR(\alpha) \\ D1(x(\alpha,TR(\alpha))) & \text{if } TR(\alpha) \le t \end{cases}$$

and again we use the function y(α,t) to determine T :

$$y(\alpha,t) := t\cdot D(\alpha,t) - A(\alpha,t)$$

(8b)

$$z(\alpha,t) := y(\alpha,Tp(\alpha)) + \frac{y(\alpha,TR(\alpha)) - y(\alpha,Tp(\alpha))}{TR(\alpha) - Tp(\alpha)}\cdot(t - Tp(\alpha))$$

(9b)

The approximation z is not as good as in the previous section but still reasonable around y = 0 .

Example :

$$t := Tp(\alpha), Tp(\alpha) + 0.01 .. TR(\alpha)$$

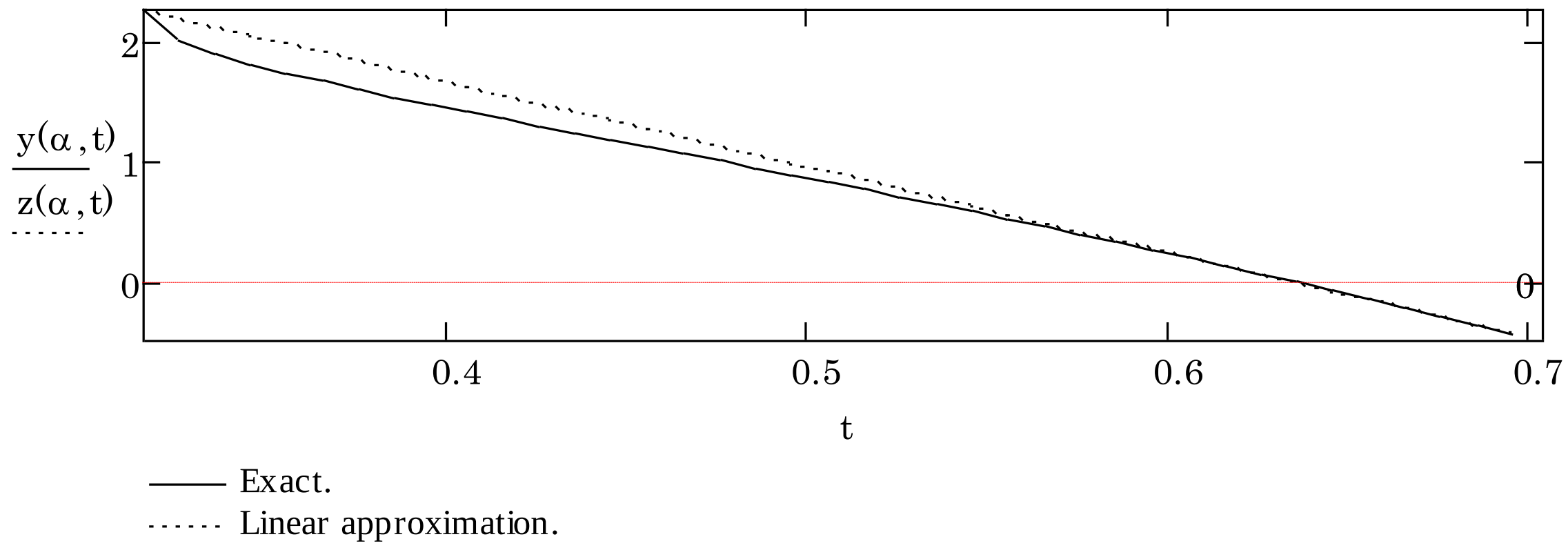

**Figure 8b. Solution for *T*.**

With the dimensionless quantity

$$\rho(\alpha) := \frac{-y(\alpha,Tp(\alpha))}{y(\alpha,TR(\alpha)) - y(\alpha,Tp(\alpha))}$$

we have

$$T(\alpha) := Tp(\alpha) + \rho(\alpha)\cdot(TR(\alpha) - Tp(\alpha))$$

and therefore

$$Ta(\alpha) := T(\alpha) - Tp(\alpha)$$

(10b)

Example :

$$meanT := \frac{1}{\pi}\cdot\int_0^{\pi} T(\alpha)\,d\alpha$$

$$meanT = 0.68$$

The mean value with respect to of **the maximum average detection width** for the example is

$$\text{meanD} := \frac{1}{\pi} \cdot \int_0^{\pi} \frac{A(\alpha, T(\alpha))}{T(\alpha)} \, d\alpha$$

meanD= 1.60

compared to the continuous case :

D1(R) = 0.91

**The Gain** G compared to the continuous case is then on the average

$$G := \begin{cases} \dfrac{\text{meanD}}{\text{D1}(R)} - 1 & \text{if } 0 < \text{D1}(R) \\ 100 & \text{otherwise} \end{cases}$$

$G = 76\cdot\%$

The mean of **the maximum average detection area per time unit** is approximate
$meanW\cdot meanD = 34$

# 8 TARGET SPEED GREATER THAN SEARCHER SPEED

We change the example to

$V := 8$

, so now $V < U$ , all other values are as previous stated.
The numerical values of W will change :

$$W(\alpha) := \sqrt{U^2 + V^2 - 2\cdot U\cdot V\cdot \cos(\alpha)}$$

and so will

$$\gamma(\alpha) := \operatorname{asin}\left(\frac{U}{W(\alpha)}\cdot \sin(\alpha)\right)$$

When $V < U$ , the only possibility for detection is when $r < S$ , as in the previous section. Going through the previous section, keeping in mind $V < U$ , and looking at figure 10, we assume an alerted target course $\alpha_e$ , corresponding to a relative radial target course away from the searcher as the optimum evasive action of the target (ideal target, worst case for the searcher). The relative target speed is called $W_e$ . The two cases, Case 1) and Case 2), still applies. The only change is the term $b/W_e$ in T1.

### 8.1 The passive period

**For $d \le r$** we have from figure 10

$$\gamma(\alpha) + \operatorname{asin}\left(\frac{d}{r}\right) + \gamma_e = \pi$$

, i.e. radial relative course away from the searcher, giving

$$\gamma_e(\alpha, d) := \pi - \gamma(\alpha) - \operatorname{asin}\left(\frac{d}{r}\right)$$

and

$$U^2 = V^2 + W_e^{\,2} - 2\cdot V\cdot W_e\cdot \cos\left(\gamma_e(\alpha, d)\right)$$

giving

$$W_e(\alpha, d) := V\cdot \cos\left(\gamma_e(\alpha, d)\right) + \sqrt{U^2 - V^2\cdot \sin\left(\gamma_e(\alpha, d)\right)^2}$$

$$T1(\alpha, d) := \frac{a(d)}{W(\alpha)} + \frac{S - r}{W_e(\alpha, d)}$$

For a(d), see Eq.(11).

**For $d \ge r$** we still have

$$T2(\alpha) := \frac{\sqrt{R^2 - S^2}}{W(\alpha)}$$

as minimum.

$$Tp\alpha d(\alpha,d) := \begin{cases} \dfrac{\sqrt{R^2-S^2}}{W(\alpha)} & \text{if } \dfrac{\sqrt{R^2-S^2}}{W(\alpha)} \le \dfrac{a(d)}{W(\alpha)} + \dfrac{S-r}{W_e(\alpha,d)} \\ \dfrac{a(d)}{W(\alpha)} + \dfrac{S-r}{W_e(\alpha,d)} & \text{otherwise} \end{cases}$$

If we have no knowledge of d, we will use the minimum of Tαd(α,d) with respect to d in order to minimize Tp in accordance with MOE1 and still keep as parameter :

$$mind(\alpha) := Minimize(Tp\alpha d, d)$$

We have

$$Tp(\alpha) := \begin{cases} \dfrac{\sqrt{R^2-S^2}}{W(\alpha)} & \text{if } \dfrac{\sqrt{R^2-S^2}}{W(\alpha)} \le \dfrac{a(mind(\alpha))}{W(\alpha)} + \dfrac{S-r}{W_e(\alpha,mind(\alpha))} \\ \dfrac{a(mind(\alpha))}{W(\alpha)} + \dfrac{S-r}{W_e(\alpha,mind(\alpha))} & \text{otherwise} \end{cases}$$

(4c)

$$meanTp := \frac{1}{\pi}\cdot\int_0^{\pi} Tp(\alpha)\,d\alpha$$

$$meanTp = 0.98$$

If we have no knowledge of we prefer the absolute minimum of Tpd (MOE1) :

$$\begin{pmatrix} \alpha minabs \\ dminabs \end{pmatrix} := Minimize(Tp\alpha d, \alpha, d)$$

, giving

$$Tp\alpha d(\alpha minabs, dminabs) = 0.41$$

## 8.2 The active period

In this section (V<U) limiting lines of approach for the target does not exist. Opportunities will only be created during the passive period in the sector S - r . As soon as the active period begins targets will be able to flee the searcher in a relative radial course away from the searcher (ideal targets) trying to avoid detection. This means that the active period should be as short as practical possible, just long enough to detect eventual targets in the sector S - r and not longer than

$$Ta(\alpha,d) := \frac{S-r}{W_e(\alpha,d)}$$

, which is the second term in T1(,d) section 8.1 above.

The absolute minimum of Ta(,d) is

$$\begin{pmatrix} \alpha min \\ dmin \end{pmatrix} := Minimize(Ta, \alpha, d)$$

, giving

$$Ta(\alpha min, dmin) = 0.03$$

# 9. EXTENSIONS

The present model is almost the simplest possible. A number of extensions can make the model more realistic :

V can be split up into speed in the passive period $V_p$ and speed in the active period $V_a$ to accommodate dipping sonar and sprint and drift tactics.

U can be split up into unalerted speed $U_u$ and alerted speed $U_a$.

Sensor degradation due to own speed . A decreasing function S(V).

S can be split up into S for the active period and a passive detection range s for the passive period.

# 10. SUMMARY

We have used five basic parameters, which all have operational significance :

For the **searcher** : S, The range where the searcher can detect the target by means of its active sensor.
V, The speed of the searcher.

For the **target** : R, The range where the target can detect the searcher, when the searcher has his sensor on. $S < R$.
r , The range where the target can detect the searcher, when the searcher has his sensor off. $r < R$.
U, The speed of the target.

We have covered the following cases :

| | | U < V | | V < U | |
|---|---|---|---|---|---|
| S < r | $R<S\frac{V}{U}$ | Section 6<br>Total period<br>Eq.(10a) | Section 5<br>Passive period:<br>Eq.(4a) | Zero<br>probability of detection | |
| | $S\frac{V}{U}<R$ | Section 6.1<br>Active period | | | |
| r < S | $R<S\frac{V}{U}$ | Section 7.2<br>Total period<br>Eq.(10b) | Section 7.1<br>Passive period<br>Eq.(4b) | Section 8.1<br>Passive period<br>Eq.(4c) | Section 8.2<br>Active period |
| | $S\frac{V}{U}<R$ | Section 6.1<br>Active period | | | |

Always S<R and r<R

Total period = Passive + Active period

**Table 1. Cases depending on five basic parameters.**

When using the sensor intermittently and from chosen values of these five parameters one can determine the passive and active period by means of our simple model . The resulting effect of the intermittent sensor use is then given as an average detection width, which is proportional to the probability of detection but has an operational significance in itself.

# Appendix A

The area A under the detection width D.

We define

$$A1(\alpha,t) = \int_{Tp(\alpha)}^{t} D1(x(\alpha,\tau))\,d\tau$$

which can be written

$$A1(\alpha,t) = \int_{r}^{x(\alpha,t)} \frac{D1(x)}{x'(\alpha)}\,dx$$

From Eq.(6a) we have

$$x'(\alpha) = \begin{cases} \dfrac{R-r}{TR(\alpha)-Tp(\alpha)} & \text{if } R < S\cdot\dfrac{V}{U} \\[2ex] \dfrac{S\cdot\dfrac{V}{U}-r}{TR(\alpha)-Tp(\alpha)} & \text{otherwise} \end{cases}$$

Using Eq.(2), we have

$$A1(\alpha,t) = \begin{cases} 2\cdot\dfrac{TR(\alpha)-Tp(\alpha)}{R-r}\cdot\left[S\cdot\sqrt{1-\left(\dfrac{U}{V}\right)^2}\cdot(x(\alpha,t)-r)-\dfrac{U}{V}\cdot\displaystyle\int_{r}^{x(\alpha,t)}\sqrt{x^2-S^2}\,dx\right] & \text{if } R < S\cdot\dfrac{V}{U} \\[2ex] 2\cdot\dfrac{TR(\alpha)-Tp(\alpha)}{S\cdot\dfrac{V}{U}-r}\cdot\left[S\cdot\sqrt{1-\left(\dfrac{U}{V}\right)^2}\cdot(x(\alpha,t)-r)-\dfrac{U}{V}\cdot\displaystyle\int_{r}^{x(\alpha,t)}\sqrt{x^2-S^2}\,dx\right] & \text{otherwise} \end{cases}$$

We define

$$I(x) = S\cdot\sqrt{1-\left(\frac{U}{V}\right)^2}\cdot x-\frac{U}{V}\cdot\int\sqrt{x^2-S^2}\,dx$$

, giving

$$I(x) := S\cdot\sqrt{1-\left(\frac{U}{V}\right)^2}\cdot x-\frac{U}{V}\cdot\frac{1}{2}\cdot\left(x\cdot\sqrt{x^2-S^2}-S^2\cdot\ln\left(x+\sqrt{x^2-S^2}\right)\right)$$

and therefore

$$A1(\alpha,t) := \begin{cases} 2\cdot\dfrac{TR(\alpha)-Tp(\alpha)}{R-r}\cdot(I(x(\alpha,t))-I(r)) & \text{if } R < S\cdot\dfrac{V}{U} \\[2ex] 2\cdot\dfrac{TR(\alpha)-Tp(\alpha)}{S\cdot\dfrac{V}{U}-r}\cdot(I(x(\alpha,t))-I(r)) & \text{otherwise} \end{cases}$$

(A1)

we have

$$A(\alpha,t) := \begin{cases} 0 & \text{if } 0 \le t \wedge t < Tp(\alpha) \\ A1(\alpha,t) & \text{if } Tp(\alpha) \le t \wedge t < TR(\alpha) \\ A1(\alpha,TR(\alpha)) + D1(R)\cdot(t-TR(\alpha)) & \text{if } TR(\alpha) \le t \end{cases}$$